\documentstyle[12pt]{article}
\begin{document}
\begin{center}{\Large {\bf Bose-Einstein Condensation in arbitrary
 dimensions}}\end{center}

\vskip 1cm

\begin{center}
{\it Ajanta Bhowal Acharyya}\\
{\it Department of Physics}\\
{\it Lady Brabourne College}\\
{\it P-1/2, Suhrawardy Avenue}\\
{\it Calcutta-700017, India}\\
{E-mail:ajanta.bhowal@gmail.com}\\
{and}\\
{\it Muktish Acharyya}\\
{\it Department of Physics}\\
{\it Presidency University}\\
{\it 86/1 College Street}\\
{\it Calcutta-700073, India}\\
{E-mail:muktish.acharyya@gmail.com}
\end{center}

\vskip 1cm

\noindent {\bf Abstract:} The density of bosonic states are calculated for
spinless free
 massive bosons in generalised {\it d} dimensions. 
The number of bosons are calculated in the lowest energy state. The
Bose Einstein condensation was found in generalised  $d$ 
dimensions (on and above 
$d = 3$)
and the condensation temperature is calculated which is
observed to drop abruptly above three dimensions and 
decreases monotonically as the dimensionalities of the system 
increases. The rate of fall of the condensation temperature decreases as
the dimensionality increases. Interestingly, in the limit $d \to \infty$,
the condensation temperature is observed to approach a nonzero finite
value.

\vskip 2cm



\newpage
Bose-Einstein condensation (BEC) \cite{huang,pathria}
is a very interesting and important
subject of modern research. 
The slowing of atoms by the use of cooling apparatus produces a 
singular quantum state known as a Bose condensate or 
Bose–Einstein condensate. 
Einstein demonstrated that cooling of 
bosonic atoms to a very low temperature 
would cause them to fall (or "condense") into the lowest accessible 
quantum state, resulting in a new form of matter.
This transition occurs below a critical temperature, which for a uniform 
three-dimensional gas consisting of non-interacting particles with no 
apparent internal degrees of freedom.
After its experimental discovery\cite{expt} the BEC becomes much more
challenging in modern research.
Very recently, the condensation of photons (massless bosons) was observed
experimentally\cite{photon} in an optical microcavity.
   
Recently, the static properties of positive ions in 
atomic BEC\cite{massigan}, the band structures, elementary excitations,
stability of BEC in periodic potentials\cite{machholm} and the
collisional relaxation in diffuse clouds of trapped bosons\cite{pethic}
are studied in physical systems. 

However, all these studies are done in three dimensions. Although, it is
a pedagogical matter, one may try to see the condensation in higher
dimensions, or more generally in $d$ dimensions. 

In this letter, we study the Bose Einstein
condensation in generalised {\it d} dimensions. 
We hope, that the present study will motivate the researchers
to do such type of generalisation of some other physical phenomena 
in higher dimensions.

In $d$-
dimensions, the density of states for non-interacting spinless bosons 
is same as that for free fermions\cite{ma1}.
 The energy of a free boson can be written as 
\cite{pathria} (Appendix-C)

\begin{equation}
E={{1} \over {2m}}(p_1^2+p_2^2+p_3^2+.....+p_d^2).
\end{equation}

\noindent Here, the above relation represents an equation of $d$-dimensional
hypersphere (in momentum space) having radius $R={\sqrt{2mE}}$. The
density of single bosonic states will be proportional
to the volume of the spherical shell 
bounded between energy $E$ and $E+dE$. This
can be calculated easily\cite{ma1}, and gives
\begin{equation}
g_d(E) dE = C(m,V) E^{{d-2} \over {2}} dE,
\end{equation}

\noindent where, the constant $C(m,V)$ is given by
\begin{equation}
C(m,V)={V\over{\Gamma(d/2)}}({{2m\pi}\over {h^2}})^{d/2}
\end{equation}

It may be readily checked that in 
three dimensions ($d=3$), one gets the well
known and widely used result of density of states,
i.e., $g(E) \sim E^{1/2}$.  

The thermodynamics of the system can be easily obtained from the 
grand canonical partition function, $Z_g(z,V,T)$, which is given by
\begin{equation}
Z_g(z,V,T)={\prod}_i{(1-ze^{-\beta E_i})}^{-1}
\end{equation}

where $\beta= 1/kT$ and $z$  is the fugacity of the gas and related to the
 chemical potential 
$\mu$ by the relation,  $z=e^{-\beta \mu} $. Here the suffix $i$ refers to 
the single particle state, having energy $E_i$ and  the product is over all
the single particle states.

The total number of particles of the
system are given by the following formula,
\begin{equation}
N={\Large {\Sigma}}_i{1\over {z^{-1}e^{\beta E_i} -1} }
\end{equation}
It is to be noted that $ze^{-\beta E}<1$ for all $E$, and since for free 
particle minimum value of $E$ is zero, the maximum value of $z$ is 1.
The number of particle $N_0$ at ground state $E=0$ is 
\begin {equation}
N_0 ={1\over {z^{-1}-1}}
\end{equation}
For large volume, the spectrum of free particle becomes almost continuous and
hence the summation in the above equation could be represented by 
$\int g_d(E)dE$. However, at $E=0$, $g_d(E)$ becomes zero (eqn (2)), hence the 
contribution of the number of particles ($N_0$) in the ground state ($E=0$)
 to the total number of particles ($N$) could not
be obtained in this way. So, one has to seperate the number $N_0$ from the
total number $N$. This trick is applied here since one finds that $N_0$
becomes considerably large as $z \to 1$.
Thus, the total number $N$ is written as
\begin{equation}
N = N_0 +{\Large {\int}}_{0^{+}}^{\infty}{{g_d(E)dE}
\over {z^{-1}e^{\beta E}-1}}.
\end{equation}

From the expression of $N_0$, it is quite clear that as $z$ 
approaches to one,
 the $E=0$ state starts to populate.
This phenomenon of accumulation of particles in the ground
state (even at $T\neq0$) is known as Bose condensation.
\begin{equation}
N-N_0 = C(m,V){\Large {\int}}_0^{\infty}{{E^{d/2 -1} dE}\over{z^{-1}e^{\beta E}-1}}
\end{equation}

\begin{equation}
N-N_0={{C(m,V)} \over {\beta^{d/2}}}
{\Large {\int}}_0^{\infty}{{X^{d/2 -1}dX} \over {z^{-1}e^{X}-1}}
\end{equation}
where $X=\beta E$
Substituting the value of $C(m,V)$ from equn. no(3), one obtains
\begin{equation}
{N-N_0\over V} = {1\over {\lambda^d}}g_{d\over2}(z)
\end{equation}
and $\lambda$ is given by
\begin{equation}
\lambda =  \sqrt{ { {2m\pi kT}\over {h^2} } }
\end{equation}
and $g_\nu(z)$ is the Bose-Einstein function, given by
\begin{equation}
g_{\nu}(z) = {1\over {\Gamma(\nu)}}{\Large{\int}}_0^{\infty}{{x^{\nu -1}}dx\over{ z^{-1}e^x -1}}
\end{equation}

Since, for bosons, the value of $z$ is always  $\le 1$, so $g_{\nu}(z)$
can be expressed as a power series,
\begin{equation}
g_{\nu}(z) = {\Sigma}_{l=1}{{z^l}\over {l^{\nu}} }
\end{equation}
The Bose condensation temperature ($T_B$) is obtained here from the equation
for $z=1$ (since $\mu=0$) and $N_0=0$. Hence $T_B$ is given by 
\cite{pathria}
\begin{equation}
{T_B}^{d\over 2}={{N/V}\over {\zeta (d/2)}}{({{h^2}\over {2m\pi k}})}^{d/2}
\end{equation}
where $\zeta (\nu ) = g_\nu (z=1)$ and is given by
\begin{equation}
\zeta (\nu) = {\Sigma}_{l=1}^{\infty}{1 \over {l^\nu}}
\end{equation}
For $\nu\le 1$, $g_\nu(z)$ diverges 
\cite{pathria}(Appendix-D) as $z \to 1$, thus for one and 
two dimensions
Bose condensation temperature($T_B$)(see eqn.(14)) is zero.
and for $d=3$  one obtains the well known result,
\begin{equation}
T_B = {({ {h^2}\over {2m\pi k} })}{{\Large(}{ {N/V}\over {2.612} }{\Large )} }^{2/3} 
\end{equation}
The condensation temperature $T_B$ can be calculated in any dimension ($d$),
from equation (14). We have estimated, the condensation temperature $T_B$,
for Helium in various dimensions and shown in Table-I.

\vskip 2cm
\begin{center}{Table-I}\end{center}
\begin{tabular}{|c|c|}
\hline
Dimensionality ($d$)& Condensation temperature ($T_B$)  \\
 & (in Kelvin)  \\
\hline
$d=3$ & 3.13269(see Ref.\cite{pathria})   \\
\hline
$d=4$ & $8.76365\times10^{-4}$   \\
\hline
$d=5$ & $5.83222\times10^{-6}$   \\
\hline
$d=6$ & $1.99971\times10^{-7}$   \\
\hline
$d=7$ & $1.77437\times10^{-8}$   \\
\hline
$d=8$ & $2.86711\times10^{-9}$   \\
\hline
$d=9$ & $6.92417\times10^{-10}$   \\
\hline
$d=10$ & $2.21788\times10^{-10}$   \\
\hline
\end{tabular}

\vskip 1cm

Here, we have used the value of $V=27.6 c.c./mole$
(see Ref.\cite{pathria}) and 
$m=6.65\times10^{-24}$. The results given in Table-I shows that the 
condensation temperature falls abruptly in $d=4$. As the dimensionalities
of the system increases, the condensation starts at lower temperature. It
is also evident from the data that the rate of fall of the condensation
temperature with respect to the dimensionalities decreases as the
dimensionality increases. Now, the question is, should we find the BEC at
zero temperature in infinite dimensions ? Interestingly, the answer is "NO".
Taking the limit $d \to \infty$ in the expression (14), we get 
$T_B(d\to \infty)$ is equal to $h^2/(2\pi m k)$\cite{referee}. 
Here, it may be noted
that $(N/(V\zeta(d/2)))^{2\over d}$ approaches unity as $d\to \infty$.

In conclusion, we can say that the present 
study of BEC in generalised {\it d}
dimensions is a new one. The condensation temperature
was found to decrease abruptly (see Table-I) in $d=4$ and then decreases
monotonically (and slowly) as the dimensionality of the system increases.
Additionally, it may also be noted (from Table-I) that the rate of fall
of condensation temperature,  
decreases as the dimensionality increases.
{\it In addition, it is quite
interesting that Bose-Einstein condensation occurs at any nonzero finite
temperature\cite{referee}, in any dimensions, 
even at $d\to\infty$, except $d=1$ and $d=2$.}

It may be mentioned here that
noninteracting electrons show some interesting unusual behaviours in higher
dimensions \cite{ma1,ma2,ma3}. In infinite dimensions, all electron
posses fermi momentum \cite{ma1}, Pauli spin susceptibility becomes
temperature independent \cite{ma2} {\it only} in two dimensions, the
form of electrical conductivity remains invariant \cite{ma3} in generalised
{\it d} dimensions. All these interesting results could not be obtained
unless studied in generalised {\it d} dimensions.

\vskip 0.5cm

\noindent {\bf Acknowledgments:} We sincerely thank the referee for suggesting
us to calculate explicitly the value of condensation temperature 
in the infinite dimensions, which constitutes the important conclusion
of this letter.

\end{document}